\documentclass[11pt,preprint]{aastex}
\usepackage{longtable}

\newcommand{\etal}{{\it et~al.}}

\begin{document}

\title{Uncertainties on Asteroid Albedos Determined by Thermal Modeling}

\author{Joseph R. Masiero\altaffilmark{1}, E.L. Wright\altaffilmark{2}, A.K. Mainzer\altaffilmark{3}}

\altaffiltext{1}{Caltech/IPAC, 1200 E California Blvd, MC 100-22, Pasadena, CA, 91125, USA, {\it jmasiero@ipac.caltech.edu}}
\altaffiltext{2}{University of California, Los Angeles, CA, 90095} 
\altaffiltext{3}{University of Arizona, Lunar and Planetary Laboratory, 1629 E University Blvd, Tucson, AZ, 85721-0092}

\begin{abstract}

We present an analysis of the accuracy of geometric albedos determined
for asteroids through the modeling of observed thermal infrared
radiation.  We show that albedo uncertainty is dominated by the
uncertainty on the measured $H_V$ absolute magnitude, and that any
analysis using albedos in a statistical application will also be
dominated by this source of uncertainty.  For all but the small
fraction of asteroids with a large amount of characterization data,
improved knowledge of the $H_V$ magnitude will be fundamentally
limited by incomplete phase curve coverage, incomplete light curve
knowledge, and the necessary conversion from the observed band to the
$V$ band.  Switching the absolute magnitude standard to a different
band such a $r'$ would mitigate the uncertainty due to band conversion
for many surveys, but this only represents a small component of the
total uncertainty.  Therefore, techniques making use of these albedos
must ensure that their uncertainties are being properly accounted for.

\end{abstract}

\section{Introduction}

Thermal infrared sky surveys have produced infrared measurements of a
large number of the known asteroids in the inner Solar system.
Application of thermal models to these data have resulted in diameter
and albedo constraints for over 100,000 asteroids from IRAS
\citep{irasPDS}, AKARI \citep{usui11}, Spitzer \citep{trilling10}, and
WISE/NEOWISE \citep{mainzer11} combined.  \citet{mainzerAIV} present an
overview of space-based studies of asteroids in the infrared,
including a discussion of the techniques for thermal modeling.

Thermal infrared observations are primarily sensitive to the size of
the asteroid observed. Once the orbit of the body is constrained, the
thermal infrared flux is directly associated to the size via the
thermal model used. Across a range of compositions and optical
reflectivities, the emissivity of asteroids is consistently very close
to a value of $\sim0.9$ \citep{lim05,vernazza12}.  As shown by
\citep{harris97}, this means that changes to the measured absolute
magnitude of an asteroid have only a minor effect on the calculated
diameter.

\citet{masiero18} presented an analysis of the accuracy of
infrared diameters using the distribution of albedos observed for
asteroid families.  They found that the uncertainty on the $H_V$
absolute magnitude is a significant component of the overall albedo
uncertainty, and dominates the albedo uncertainty for typical $H_V$
accuracies.  This means that the knowledge of the absolute magnitude
plays a critical role in our understanding of albedos.

\citet{pravec12} performed an analysis of the $H_V$ values published
in asteroid orbital catalogs (the most common source for these values)
compared to objects tracked over long periods of time with
photometrically-calibrated systems.  They showed that while $H_V$
values found in orbit catalogs are generally good to a few-tenths of a
magnitude for large and well-studied asteroids, smaller objects can
show significant errors, both random and systematic, at the level of
$0.5~$mag.  This is a result of a combination of effects, including
using as assumed value for the phase curve $G_V$ parameter
\citep{bowell89}, the absolute photometric calibration of the surveys
providing photometry, some surveys using unfiltered observations for
photometry, the accuracy of the estimation of mean brightness (to
account for rotational light curve effects), changing viewing aspects
resulting in different views of the asteroid's 3D shape, and the
accuracy of the conversion from the observed band to the `standard'
$V$ band used for $H_V$ calculation.  For each of these uncertainty
components, observations can reduce their individual contribution
\citep[such as was done by][]{pravec12}, however this requires large
amounts of telescope time for each object to densely sample the
rotational light curve, phase curve, and different apparitions to
constrain the 3D shape.  The majority of objects in Minor Planet
Center's orbital catalog do not, and will not, have this level of
knowledge without targeted densely-sampled followup covering a broad
range of phase angles to constrain both the light curve and phase
curve.

Here we investigate the effect that uncertainties on $H_V$ will have
on the albedos derived when a survey provides diameter fits, as occurs
with thermal infrared data.  This is important to help us better
understand the limitations of the derived albedo data sets, and how
these uncertainties will affect our interpretations of the population
as a whole and the sub-populations within the asteroids.

\section{Relationship between diameter, absolute magnitude, and albedo}

The empirical relationship between the size of a body, its geometric
albedo, and the brightness is often described \citep[e.g.][]{harris02}
as:

\begin{equation}
D_{km} = C_V~\frac{10^{-H_V/5}}{\sqrt{p_V}}
\label{eq.orig}
\end{equation}

\noindent which can be rearranged as:

\begin{equation}
p_V = C_V^2~\frac{10^{-H_V/2.5}}{D_{km}^2}
\label{eq.alb}
\end{equation}

\noindent where $D$ is the size of the body in kilometers, $H_V$ is
the phase- and distance-corrected magnitude (i.e. absolute magnitude)
in the $V$ band, and $p_V$ is the geometric albedo in the $V$ band.
The constant parameter $C_V$ is usually taken to be $C_V=1329~$km
\citep[for example, see the derivation in][]{pravec07}.

The definition of geometric albedo is the ratio of the true scattering
of light by the surface compared to an ideal scatterer, here a disk of
area $\pi r^2 = \pi \frac{D^2}{4}$ that is 1 AU from the sun, 1 AU
from the observer, and at phase of $\alpha=0^\circ$.  Following
\citep{jewitt13}, this relationship can be written:

\[\pi \frac{D^2}{4} = (1.496e8~km)^2 \frac{\pi}{p_V}~10^{0.4(V_\sun - H_V)}\]


\[D = 2.99e8~km~\frac{1}{\sqrt{p_V}}~10^{0.2 V_\sun} 10^{-0.2 H_V}\]

\noindent meaning the relationship constant $C_V$ is:

\[C_V = 2.99e8\times10^{0.2 V_\sun}~km\]

The constant of interest is a function of the stellar apparent
magnitude of the Sun in the band of interest, here $V_\sun$ band.
\citet{torres10} quote a value of $V_\sun = -26.76 \pm 0.03$, which
they derive by recomputing the calibrations of \citet{bessell98} using
updated reference stars.  From this, we then derive a constant of
$C_V=1330\pm18~$km.  This implies that an albedo derived from a
measured diameter and an $H_V$ magnitude will automatically have a
$\sim2.8\%$ relative uncertainty from the uncertainty on the Solar V
apparent magnitude, even before accounting for errors on $D$ and
$H_V$.  The previously-derived $C_V=1329$ value, based on a
  $V_\sun=-26.762\pm0.017~$mag from \citet{campins85}, is within
measurement uncertainties of the value that is obtained with current
Solar magnitude measurements.

An important point is that this constant is a function of the bandpass
being used.  The majority of current and planned sky surveys do not
use the Bessell V filter, instead having moved to a filter set similar
to that of the Sloan Digital Sky Survey
\citep[SDSS,][]{gunn98,smith02}.  The conversion from the survey band
to $V$ will add an additional component of systematic uncertainty to
any albedos determined. Further, this conversion will depend on the
(unknown) composition of the object, as asteroids with different
spectral curves have different colors, meaning that the systematic
uncertainty will be different for different classes of object making
comparisons between populations more difficult.

One option to reduce this conversion error is for the community to
transition asteroid absolute magnitudes and albedos to a band that
dominates the ongoing survey photometry.  For example, the $r'$ band
is typically the most sensitive to asteroids for ground-based surveys
given their intrinsic brightness combined with filter responsivities.
Using $r'_\sun=-27.05\pm0.03$ \citep[Vega mags,][]{willmer18} leads to a new
constant value of $C_{r'}=1164\pm16$.  This would, or course, require
determination of the $H_{r'}$ absolute magnitude and the
$G_{r'}$ phase parameter for all asteroids being studied, as well as a
conversion technique to compare new albedos to literature $p_V$
values.  However, as surveys such as the Legacy Survey of Space and
Time at the Vera Rubin Observatory \citep{lsst} begin producing large
quantities of asteroid photometry over many years, these measurements
will become possible.  Given that the majority of asteroid photometry
over the next decade will be obtained in $r'$ or a closely calibrated
band, it is worth careful consideration by the community whether now
is the time to switch standards.

In counterpoint, there are arguments against switching standards as
well.  Foremost is the extensive amount of literature currently using
$H_V$ and $p_V$, as well as the numerous diagnostics that exist based
on these parameters.  In addition, the $V$ band covers the peak of the
distribution of reflected light from an asteroid. That makes it a
closer analog to the true bolometric albedo, which is an important
value needed for thermophysical modeling. Instead of switching
standards, a concerted effort to provide accurate $V-r'$ indices for
all surveys for a range of asteroid compositions might alleviate some
of the problems created by the current system.  Any change, of course,
would require extensive community discussion and IAU approval.

\section{Absolute magnitude uncertainty} 

Following Eq~\ref{eq.alb}, we can see that the error on albedo will be
a combination of the errors on diameter and absolute magnitude.  While
diameter error can be independently assessed based on comparisons
between different determination methods (e.g. infrared modeling, radar
modeling, or occultation chord fits), the true uncertainty on $H_V$ is
more difficult to validate against an independent dataset.

Following \citet{bowell89}, the absolute magnitude can be determined
from fitting the phase-magnitude relationship of the asteroid using
the equations:

\begin{equation}  
  H_V = V_{obs} + 2.5 \log_{10}((1-G_V)~\Phi_1 + G_V~\Phi_2) - 5 \log_{10}(R \Delta)
  \label{eq.HG}
\end{equation}
\[\Phi_1=\exp{\left(-3.33 \tan^{0.63}(0.5 \alpha)\right)}\]
\[\Phi_2=\exp{\left(-1.87 \tan^{1.22}(0.5 \alpha)\right)}\]

\noindent where $\alpha$ is the phase angle, $R$ is the heliocentric
distance, and $\Delta$ is the geocentric distance at the time of
observation.  This is the simplified functional form adopted by the
IAU \citep{marsden85}, though a more precise calculation is presented
by \citep{bowell89} in their equation A4.

Other photometric phase functions have been developed, such as the
H-G$_1$-G$_2$ system and the H-G$_{12}$ system \citep{muinonen10},
however as these either have more parameters (in the case of the
H-G$_1$-G$_2$ system) or non-linear behavior (in the case of the
H-G$_{12}$ system) they require more data to accurately fit the phase
curve and thus will have comparable or lower accuracy for sparse data
sets.

We note that in well-defined cases with extensive data and
multi-parameter fits, like those presented in \citet{muinonen10}, the
uncertainty on $H_V$ can be of order $0.02~$mag (1-$\sigma$). However
as the authors of that paper note, a number of factors commonly
encountered with photometric data can impair the determination of
$H_V$ including changing geometry between apparitions, incomplete
rotational coverage at each phase angle, coverage of only a narrow
range of phase angles, and imperfect conversion of photometry from the
observed band to $V$.

The work of \citet{veres15} provides an ideal example of real world
results from fitting absolute magnitudes to a large,
photometrically-calibrated survey that is sparse in time.  In that
work, the vast majority of observations had individual photometric
uncertainties $<0.1~$mag, meaning that the individual observations did
not place a fundamental limit on the accuracy of the $H_V$
determination.  Through Monte Carlo simulations of different rotation
states, those authors found that their statistical uncertainty on
$H_V$ was $0.3~$mag for objects with $H<18~$mag (sizes larger than
approximately 1 km) using the H-G relation from \citet{bowell89}, or a
slightly improved uncertainty of $0.25~$mag under the H-G$_{12}$
system developed by \citep{muinonen10}.  For cases where diameters are
measured with an infrared survey with a nominal accuracy of $10\%$,
the above uncertainty on $H_V$ would result in a relative uncertainty
on albedo of $32-36\%$.  In the converse case of an object with only
optical observations and using an assumed albedo with perfect
accuracy, this $H_V$ uncertainty alone would propagate to an
uncertainty on diameter of $13-15\%$, with additional non-random
uncertainty from the accuracy of the assumed albedo used.

\citet{veres15} note that the fits to the $G_V$ slope parameter are
significantly worse than the $H_V$ fits in their work.  The
uncertainty on $G_V$ depends on the span of phase angles covered, with
coverages $>20^\circ$ showing significantly smaller uncertainties than
those with coverage $<10^\circ$.  In addition, objects that are only
seen at high phase angles will have significant errors on $H_V$ due to
the large lever-arm that the fitted value of $G_V$ has, as is often
the case for near-Earth asteroids.  Due to this uncertainty on $G_V$,
$H_V$ errors of $\sim1~$mag can be expected for objects on Earth-like
orbits.  This level of uncertainty for the $H_V$ value would
correspond to an albedo uncertainty of $\sim70\%$ when using a
infrared-determined diameter.  In the case of a diameter calculated
from $H_V$ and an assumed albedo, the uncertainty on this diameter
will be $\sim42\%$. As $G_V$ is only weakly correlated with taxonomy
\citep[see Table 6 from][]{veres15}, asteroid color measurements cannot
dramatically improve this.

\begin{figure}[ht]
\begin{center}
  \includegraphics[scale=0.8]{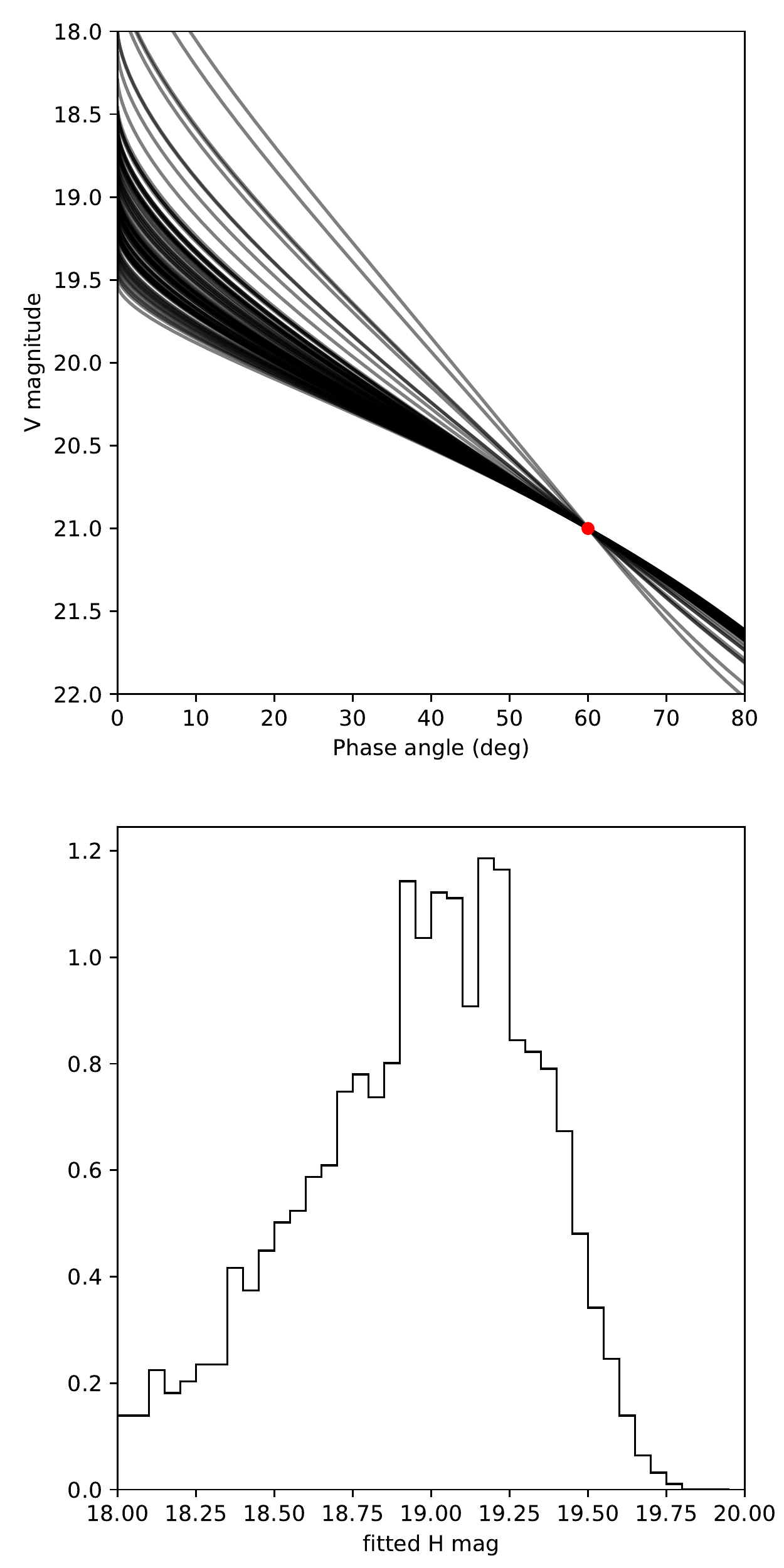} \protect\caption{An
    example of the range of allowable phase curve fits for an asteroid
    detected only at high phase angles. (Top) The red circle shows the
    simulated detection, with possible H-G phase relations shown as
    gray lines.  (Bottom) Normalized histogram of possible $H_V$
    values in this example for 2000 Monte Carlo simulations of the
    $G_V$ parameter.}
\label{fig.HGfit}
\end{center}
\end{figure}

An example of this situation is shown in Fig~\ref{fig.HGfit}.  Here,
we assume an asteroid is detected with $V=21~$mag (with negligible
measurement uncertainty) at a phase of $\alpha=60^\circ$, 1~AU from
the Earth and Sun.  Based on \citep{veres15} we drew random $G_V$
parameters from a normal distribution following $G_V=0.2\pm0.2$.  This
results in a median absolute magnitude of $19.00^{+0.29}_{-0.54}$
(uncertainties shown are the 84$^{th}$ and 16$^{th}$ percentiles).
For objects typically observed at smaller phase angles, like Main Belt
asteroids, this situation will not occur and $H_V$ values will be
better constrained.  However, near-Earth objects are often observed at
large phase angles, where this problem can result in significant
uncertainties in their absolute magnitudes. For example, cadence
simulations for the Vera C. Rubin Observatory's Legacy Survey of Space
and Time (LSST) indicate that the twilight NEO survey, if carried out,
would have approximately half of the survey fields at Solar
elongations below $110^\circ$, resulting in NEO detections
predominantly at large phase angles \citep{jones20}.

As the range of phase angles covered increases, the constraint on the
$G_V$ value for an object will improve, and thus the $H_V$
determination will also improve, however this is a strong function of
the phase angles at which the object is observed.  A Monte Carlo
simulation of four $V$-band observations (assuming zero photometric
error and perfect accuracy on the assumed albedo) in the range
$10^\circ<\alpha<30^\circ$ yields a final uncertainty on $H_V$ of
$0.23~$mag and minimum calculated diameter uncertainty of $11\%$,
while four observations drawn from $50^\circ<\alpha<70^\circ$ result
in an uncertainty of $0.68~$mag on $H_V$ and $30\%$ on $D$.
Increasing the number of observations to $64$ further improves the
$H_V$ uncertainty to $0.08~$mag ($4\%$ on $D$) for $10-30^\circ$ and
$0.31~$mag ($14\%$ on $D$) for $50-70^\circ$.  Photometric measurement
uncertainty will increase the true uncertainty on both $H_V$ and $D$,
while the uncertainty on the assumed albedo will increase diameter
uncertainty as well.

\clearpage

It should be noted that these results depend on the assumption that
the H-G or H-G$_{12}$ phase functions can adequately describe the
opposition effect of all objects.  Recent work by \citet{mahlke20}
investigated the photometric phase curves of over 90,000 asteroids in
two visible light bandpasses from the ATLAS survey.  They show that
even for objects with well-sampled phase curves, the difference in
fitted absolute magnitude between the H-G$_{12}$ system and the
H-G$_{1}$-G$_{2}$ has systematic offsets of order $0.1~$mag and random
uncertainties of $\sim0.2~$mag. This will directly impact the accuracy
of the albedos determined for asteroids from thermal modeling.

\section{Discussion}

The overall uncertainty on any given asteroid's albedo measurement is
a combination of the uncertainties on the $H_V$ value and the diameter
derived from thermal modeling.  Comparisons of diameters determined
from thermal modeling to those from different surveys as well as
independent sources such as occultations and radar observations have
shown that when multiple infrared measurements are available that
sample the thermal emission portion of an asteroid's spectral energy
distribution, the diameter uncertainty is approximately $10\%$
\citep{mainzer11cal,usui14,wright18,herald20}.  This uncertainty is primarily
caused by the deviation of the applied thermal model (usually a sphere
with a simple temperature profile) from the asteroid's actual
thermophysical properties.  This propagates to an albedo uncertainty
of $\sim20\%$, which is comparable to the albedo uncertainty resulting
from a typical $H_V$ uncertainty.  Along with the uncertainty on the
constant in Eq~\ref{eq.alb} the result is a top-level albedo
uncertainty of $\sim28\%$ for typical values of well-studied objects
of $\sigma_D=10\%$ and $\sigma_{H_V}=0.2~$mag.

Using large samples of objects with assumed uniform properties,
e.g. from asteroid families \citep{masieroAIV} or selected by
photometric colors \citep{ivezic20}, it is possible to obtain a mean
albedo for the population that is known to higher precision than any
single object's albedo.  In this case, the uncertainty on a diameter
inferred using this assumed albedo would be dominated by the accuracy
with which the object has been assigned an albedo and the uncertainty
on $H_V$.  For a case of $\sigma_{H_V}=0.2~$mag and an albedo assumed
to be known with arbitrary precision, the uncertainty on diameter will
be $\sim10\%$.  As shown by \citet{pravec12}, $\sim0.2~mag$ is the
smallest $H_V$ uncertainty that can be reasonably expected from
current data for objects with $D<10~$km, even after correcting for
systematic errors in the orbital catalogs.  For objects that are newly
discovered, light curve properties and phase curve behavior will not
be characterized to high precision without many years of observations,
and the characterization accuracy will depend on the the observing
cadence.  These objects will thus have commensurately worse $H_V$
constraints, which would translate to larger uncertainties on inferred
diameter even under the assumption of perfect albedo assignment.  As
discussed above, for newly discovered NEOs the diameter uncertainty
from optical data alone can reach $\sim50\%$, particularly for objects
on Earth-like orbits.

\section{Conclusions}

As is noted by both \citet{bowell89} and \citet{muinonen10}, the
geometric albedo determined from the relation in Eq~\ref{eq.orig}
might be more appropriately called a `pseudo-albedo' as it is not a
direct measurement, but rather inferred from models of models of
measurements.  This is not to say that it is not useful for population
analysis or investigations of individual objects, as this is clearly
demonstrated in the literature.  Rather, as highlighted by
\citet{bowell89}, this relationship should be treated with caution as
multiple assumptions go into a single derived value.  As we have
discussed here, the uncertainty on $H_V$ can have a large impact on
our knowledge of $p_V$, and it is nearly impossible to independently
verify $H_V$ measurements against other, non-photometric data sources.
In light of this, we urge caution in attempting to derive physical
properties from albedos alone.

\textbf{Acknowledgments:} We thank the referees Petr Pravec and Alan
Harris for their helpful reviews that improved this paper.  This
research has made use of NASA’s Astrophysics Data System.

\clearpage

\end{document}